\documentclass[conference]{IEEEtran}
\IEEEoverridecommandlockouts

\usepackage{amsmath,amssymb}
\usepackage{graphicx}
\usepackage{textcomp}
\usepackage{xcolor}
\usepackage{cite}
\usepackage{float}
\usepackage{subcaption}  
\usepackage{algorithm}
\usepackage{algpseudocode}
\usepackage{stfloats}
\usepackage{booktabs}
\usepackage{tabularx}
\usepackage{hyperref}
\usepackage{amssymb} 
\usepackage{bm}
\usepackage{comment}

\begin{document}

\title{Joint Electricity Price and Load Forecasting for Data Center Battery Dispatch Using Polynomial Diffusion Models

\thanks{This work was supported by the Natural Sciences and Engineering Research Council (NSERC) Discovery Grant, NSERC RGPIN-2022-03236 and by Fonds de recherche du Québec under Grant FRQ-NT PR-378486 (https://doi.org/10.69777/378486).}
}

\author{Matthew~Van~Ham~\IEEEmembership{Student Member}, Yiping Liu~\IEEEmembership{Student Member}, Xiaozhe Wang~\IEEEmembership{Senior Member}}%

\maketitle

\begin{abstract}
The rapid growth of large-scale data centers (DCs) is increasing electricity demand, market volatility, and operating costs. Battery energy storage systems (BESSs) can mitigate these challenges by shifting electricity purchases from high-price to low-price periods. This paper proposes a real-time BESS dispatch framework using polynomial diffusion models (PDMs) to forecast both electricity prices and DC loads. Leveraging the closed-form conditional expectations of PDMs, the forecasts are integrated into a dynamic programming (DP) framework to minimize electricity procurement costs. Across the evaluated BESS configurations, the proposed approach achieves 2.63 times the cost savings of a rule-based (RB) strategy.
\end{abstract}


\section{Introduction}

The rapid global expansion of data centers (DCs), driven by artificial intelligence and cloud computing, is increasing pressure on electricity infrastructure and operating costs \cite{unece2026ai}. Efficient electricity procurement and demand management are therefore critical to supporting DC growth while mitigating impacts on power systems \cite{wierman2014opportunities}. Battery energy storage systems (BESSs) offer operational flexibility by shifting electricity purchases from high-price to low-price periods.

BESS-based DC energy management was first investigated by Urgaonkar et al. \cite{urgaonkar2011} using Lyapunov optimization and subsequently extended to geographically distributed DCs \cite{guo2013}. Beyond Lyapunov-based methods, dynamic programming (DP) has been applied to battery arbitrage under stochastic electricity prices \cite{zheng2022arbitraging,diaz2018maximum}, while stochastic and distributionally robust optimization have been developed to address electricity procurement uncertainty \cite{jordehi2022risk, dabbagh2019shaving}. 
However, these studies primarily focus on electricity market uncertainty or general energy storage applications. For DCs, both electricity prices and power demand fluctuate over time, jointly influencing the profitability of battery operation. Efficiently incorporating their stochastic dynamics into real-time dispatch decisions, while characterizing the conditional distribution of cost savings, remains challenging.

This paper develops a polynomial diffusion model (PDM)-based forecasting and DP framework for cost-effective DC BESS operation. Building on the application of PDMs to electricity prices \cite{ware2019polynomial}, we make two main contributions: (i) extending PDMs to DC load modeling, enabling efficient conditional forecasts of both electricity prices and loads with online state updates; and (ii) integrating these forecasts into a receding-horizon DP framework for BESS dispatch, with Monte Carlo simulation used to characterize the distribution of realized cost savings. The proposed framework provides DC operators with a computationally tractable approach to managing electricity costs amid growing demand.

The remainder of this paper is organized as follows. Section II develops the PDM-based price and load forecasting models. Section III presents the DP-based BESS dispatch framework. Section IV validates the models and evaluates dispatch performance, while Section V concludes the paper.

\section{Model Formulation}
This section develops PDMs for electricity price and DC load forecasting. The resulting conditional expectations are used for BESS dispatch in Section \ref{sec:battery}.

\subsection{Polynomial Diffusion Models}

Electricity prices exhibit strong periodic patterns, mean reversion, volatility, and occasional extreme spikes,  
while DC loads exhibit temporal dependence and operating variability. PDMs capture these stochastic dynamics while enabling efficient computation of conditional expectations \cite{ware2019polynomial}\cite{filipovic2016}.


A polynomial diffusion is a Markov process $X_t$ on $E\subseteq\mathbb R^M$ whose infinitesimal generator $\mathcal G$, describing the instantaneous expected rate of change of a function of $X_t$, maps polynomials of degree at most $n$ into polynomials of the same or lower degree \cite{filipovic2016}. This property enables conditional expectations of polynomial functions to be computed through finite-dimensional matrix operations.

\begin{figure}
    \centering
    \includegraphics[width=0.9\linewidth]{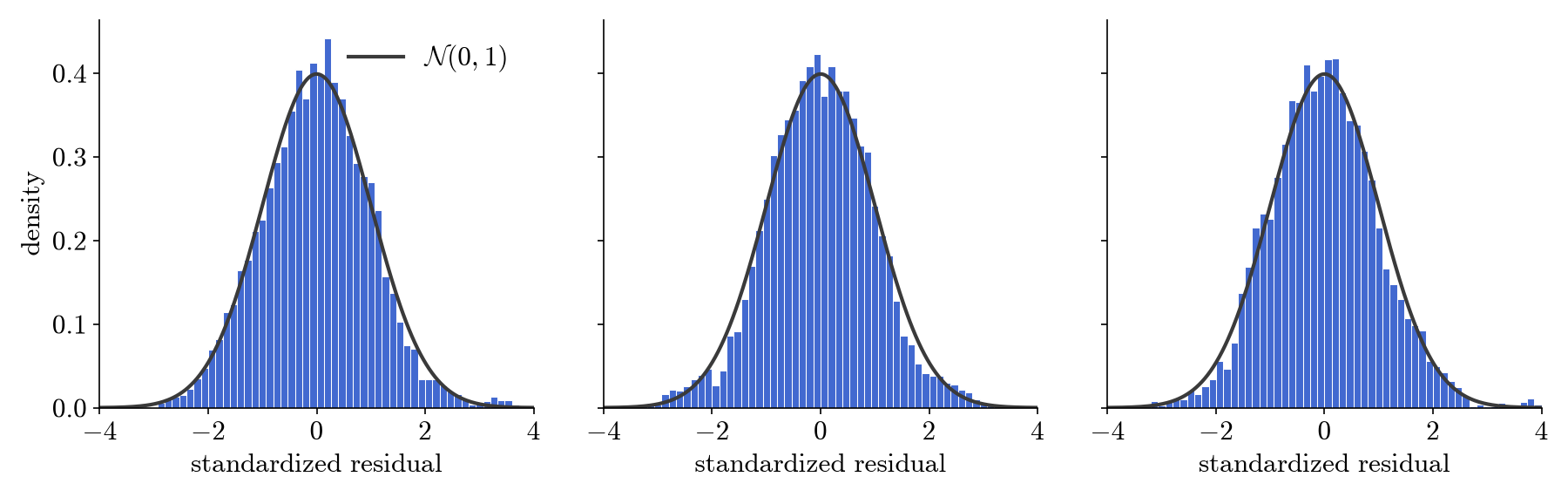}
    \caption{
    Deseasonalized power load residual for three data center series over 8,928 5-minute observations \cite{wilkes2020clusterdata}.} 
    \label{fig:loadres}
    \vspace{-0.3cm}
\end{figure}

PDMs have been successfully applied to electricity price modeling \cite{ware2019polynomial}. Here, we extend the models to DC loads, a new application investigated in this work. After removing deterministic periodic components, the residuals of DC loads exhibit approximately Gaussian behavior (Fig.~\ref{fig:loadres}). This suggests that relatively low-order polynomial transformations of mean-reverting latent factors may adequately represent load variations, whereas more strongly nonlinear transformations may be needed to capture extreme electricity price fluctuations.

Therefore, we employ PDMs for both electricity prices and DC loads within a unified stochastic framework, providing efficiently updated conditional forecasts for DP-based BESS dispatch. 

\subsection{Electricity Price and Load Representation}

In intraday electricity markets, electricity is traded for fixed delivery intervals, typically 15 minutes. 
Each contract $k$ corresponds to a delivery window $[T_k, T_k+\Delta t]$. 
We model both electricity prices and DC loads as the sum of a deterministic seasonal component and a stochastic residual. 
Specifically, the conditional interval-averaged forecasts at time \(t\) are 
\begin{equation}
\small
F_k(t)=\Phi_k^{p}(t)+\boldsymbol{\kappa}_p^\top\bar{\mathbf F}_{D,k}
\label{eq:price-decomp-elec}
\end{equation}
\begin{equation}
L_k(t)=\Phi_k^{l}(t)+\boldsymbol{\kappa}_l^\top\bar{\mathbf F}_{D,k}
\label{eq:price-decomp-load}
\end{equation}
where $\boldsymbol{\kappa}_p$ and $\boldsymbol{\kappa}_l$ are the corresponding seasonal coefficients, $\bar{\mathbf F}_{D,k}
=\frac{1}{\Delta t}\int_{T_k}^{T_k+\Delta t}
\mathbf F_D(u)\,du$ is the Fourier basis averaged over interval \(k\), 
and $\Phi_k^{p}(t)$ and $\Phi_k^{l}(t)$ are the conditional expectations of the deseasonalized price and load, respectively, obtained from their PDMs.
The seasonal components and parameters of the PDMs are calibrated offline, whereas the stochastic expectations $\Phi_k^{p}(t)$ and $\Phi_k^{l}(t)$ are updated online as new market and load observations become available.

\subsection{Seasonal Component}
\label{sec:seasonal}

Daily and weekly patterns of electricity prices and DC loads are represented by a truncated Fourier series containing daily and weekly harmonics:
\begin{equation}
\vspace{-0.1cm}
\footnotesize
\begin{aligned}
s_a(t)=&\boldsymbol\kappa_a^\top\mathbf F_D(t)={}c_{0,a}
+\sum_{j=1}^{J_1}
\left[
a_{j,a}\sin(2\pi jt)
+b_{j,a}\cos(2\pi jt)
\right]\\
&+\sum_{j=1}^{J_2}
\left[
d_{j,a}\sin\left(\frac{2\pi jt}{7}\right)
+e_{j,a}\cos\left(\frac{2\pi jt}{7}\right)
\right],
\end{aligned}
\end{equation}
where $a\in\{p,l\}$ identifies the price or load model, $t$ is measured in days, and $J_1$ and $J_2$ determine the number of daily and weekly harmonics. In this work, $J_1=J_2=5$. For delivery interval $k$, the seasonal contribution is
\begin{equation}
\bar s_{a,k}=
\frac{1}{\Delta t}
\int_{T_k}^{T_k+\Delta t}s_a(u)du
=\boldsymbol\kappa_a^\top\overline{\mathbf F}_{D,k}
\label{eq:window-ave}
\end{equation}
The seasonal coefficients $\boldsymbol\kappa_a = \{c_{0,a}, a_{j,a}, b_{j,a}, d_{j,a}, e_{j,a}\}$ are estimated by least squares using historical observations. 
Removing the estimated seasonal component yields the residual series used to calibrate the stochastic models.

\subsection{Latent Factor Dynamics}
The stochastic residuals are modeled using mean-reverting latent factors, which represent unobserved drivers of price and load variations. 
For each model $a\in\{p,l\}$, we consider   
$M_a \in \{1,2,3\}$  correlated Ornstein--Uhlenbeck (OU) factors  $\mathbf X_t^a=(X_{1,t}^a,\ldots,X_{M_a,t}^a)^\top$, governed by
\begin{eqnarray}
dX_{i,t}^a&=&-\theta_{i,a}X_{i,t}^adt+dW_{i,t}^a, \nonumber\\
d\langle W_i^a,W_j^a\rangle_t
&=&\rho_{ij,a}dt,
\qquad\rho_{ii,a}=1, \label{eq:ou}
\end{eqnarray}
where $\theta_{i,a}>0$ is the mean-reversion rate of factor $i$ and $\rho_{ij,a}$ characterizes the correlation between factors within model $a$. Different mean-reversion rates capture fluctuations across multiple time scales. 

For a discrete interval $h$, the OU dynamics admit the exact Gaussian transition
\begin{equation}
\mathbf X_{t+h}^a\mid\mathbf X_t^a
\sim\mathcal N
\left(\mathbf A_a\mathbf X_t^a,\mathbf Q_{h,a}\right),
\label{eq:exact-step}
\end{equation}
where $\mathbf A_a=
\operatorname{diag}
\left(e^{-\theta_{1,a}h},\ldots,
e^{-\theta_{M_a,a}h}\right)$ and $(Q_{h,a})_{i,j}=\rho_{i,j,a}
\frac{1-e^{-(\theta_{i,a}+\theta_{j,a})h}}
{\theta_{i,a}+\theta_{j,a}}$. 
The exact Gaussian transition provides analytical expressions for the conditional mean and covariance of future states, enabling efficient prediction and sampling without numerical discretization.  

\subsection{Polynomial Observation Mapping}
\label{sec:pricemap}
Electricity prices and DC loads may exhibit non-Gaussian fluctuations such as skew and near-discontinuous price spikes. We therefore follow \cite{ware2019polynomial} and apply polynomial observation mappings to transform the Gaussian latent factors into non-Gaussian deseasonalized price and load processes. For each model $a\in\{p,l\}$, the stochastic residual is represented as
\begin{equation}
Z_t^a=g_a(\mathbf X_t^a)=
\Psi_a\left(
\boldsymbol\alpha_a^\top\mathbf X_t^a+\beta_a
\right), \label{eq:polynomial mapping}
\end{equation}
where $\boldsymbol\alpha_a$ contains the factor loadings, $\beta_a$ is an offset, and $\Psi_a$ is an increasing odd-degree polynomial: 
\begin{equation}
\Psi_a(z)=c_{1,a}
+c_{2,a}\int_0^z
\prod_{i=1}^{r_a}p_{i,a}(y)dy,
\label{eq:psi}
\end{equation}
with
\begin{equation}
\begin{aligned}
p_{i,a}(y)={}&
\sin^2(\chi_{i,a})
(y\cos\omega_{i,a}-\sin\omega_{i,a})^2\\
&+\cos^2(\chi_{i,a})\geq0.
\end{aligned}
\end{equation}
Here $r_a$ controls the polynomial degree $D_a=2r_a+1$, while $\omega_{i,a}$ and $\chi_{i,a}$ determine the polynomial shape. The coefficients $c_{1,a}$ and $c_{2,a}$ are fixed by the normalization conditions $\Psi_a(0)=0$ and $\Psi_a(1)=1$.

The polynomial mapping transforms Gaussian latent factors into potentially non-Gaussian processes, providing the flexibility to capture asymmetric fluctuations and extreme events. 
Its monotonicity preserves the ordering of the underlying scalar latent signal, while its polynomial structure enables efficient computation of conditional expectations \cite{ware2019polynomial}\cite{filipovic2016}. These advantages also motivate its application to DC loads, whose comparatively mild stochastic fluctuations may be represented using lower-order polynomial mappings. 

\subsection{Closed-Form Conditional Expectations} 
\label{sec:forward}
The key computational advantage of PDMs is that conditional expectations of polynomial functions can be evaluated through finite-dimensional matrix operations.
For each model $a\in\{p,l\}$, the observation function $g_a$ defined in \eqref{eq:polynomial mapping} is a polynomial of degree $D_a$, which can be expressed as: 
\begin{equation}
g_a(\mathbf x)=\mathbf H_a(\mathbf x)^\top\mathbf q_a,
\end{equation}
where $\mathbf H_a(\mathbf x)$ is the vector of monomials of degree at most $D_a$, and $\mathbf q_a$ contains their coefficients.
For the OU dynamics in \eqref{eq:ou}, the infinitesimal generator is
\begin{equation}
\small
\mathcal G_a f(\mathbf x)
=-\sum_{i=1}^{M_a}\theta_{i,a}x_i\partial_i f(\mathbf x)
+\frac12\sum_{i,j=1}^{M_a}
\rho_{ij,a}\partial_i\partial_j f(\mathbf x).
\end{equation}
\normalsize
Since $\mathcal G_a$ maps polynomials of degree at most $D_a$ into the same polynomial space, its action can be represented by a matrix $\mathbf G_a$:
\begin{equation}
\small
\mathcal G_a\left(\mathbf H_a(\mathbf x)^\top\mathbf q\right)
=\mathbf H_a(\mathbf x)^\top\mathbf G_a\mathbf q.
\end{equation}
\normalsize
$\mathbf G_a$ is the matrix representation of the generator, determined by the OU parameters in \eqref{eq:ou}. Consequently, the conditional expectation of $g_a$ at horizon $s$ is
\begin{equation}
\mathbb E[g_a(\mathbf X_{t+s}^a)\mid\mathbf X_t^a=\mathbf x]
=\mathbf H_a(\mathbf x)^\top e^{s\mathbf G_a}\mathbf q_a\label{eq:gen}
\end{equation}

For contract $k$ evaluated at time $t$, define $\tau_1=T_k-t$ and $\tau_2=T_k+\Delta t-t$. The deseasonalized forward expectation $\Phi_k^a(t)$ is obtained by averaging \eqref{eq:gen} over the delivery interval:
\begin{equation}
\small
\begin{aligned}
\Phi_k^a(t)
&=\frac{1}{\tau_2-\tau_1}
\int_{\tau_1}^{\tau_2}
\mathbb E\left[
g_a(\mathbf X_{t+s}^a)
\mid \mathbf X_t^a=\mathbf x_t^a
\right]ds\\
&=\mathbf H_a(\mathbf x_t^a)^\top
\mathbf V_a
\operatorname{diag}
\left(
\varphi(\lambda_{a,i};\tau_1,\tau_2)
\right)
\mathbf V_a^{-1}\mathbf q_a,
\label{eq:conditionalex} 
\end{aligned}
\end{equation}
\normalsize
where $\mathbf G_a=\mathbf V_a\operatorname{diag}(\boldsymbol\lambda_a)\mathbf V_a^{-1}$ is an eigendecomposition of the generator matrix, and
\begin{equation}
\small
\varphi(\lambda;\tau_1,\tau_2)=
\begin{cases}
\dfrac{e^{\lambda\tau_2}-e^{\lambda\tau_1}}{\lambda(\tau_2-\tau_1)}, & \lambda\neq0,\\[1ex]
1, & \lambda=0.
\end{cases}
\end{equation}
\normalsize
Equation \eqref{eq:conditionalex} provides the deseasonalized conditional expectation for either electricity price or load. Adding the corresponding seasonal component in \eqref{eq:price-decomp-elec} or \eqref{eq:price-decomp-load} 
yields the delivery-interval forecast. Since the generator matrix and its eigendecomposition are computed offline, online forecasting requires only evaluating the polynomial basis at the updated latent state and performing matrix–vector operations, rather than stochastic simulation. 

\subsection{Parameter Estimation and Online Updating} 
\label{sec:estimation}
The unobserved latent states are estimated using an unscented Kalman filter (UKF). 
For the electricity price model, the observed quotation $Y_k^p(t_n)$ for contract $k$ at time $t_n$ satisfies 
\begin{equation}
\small
Y_k^p(t_n)=F_k(t_n;\mathbf X_{t_n}^p)
+\varepsilon_{k,n}^p
\label{eq:observation}
\end{equation}
\normalsize
where 
$\varepsilon_{k,n}^p
\sim\mathcal N
\left(0,\sigma_{\varepsilon,p}^2\tau_{k,n}\right)$ and $\tau_{k,n}=T_k-t_n$ denotes the time to delivery. 
Stacking the $d_n$ contracts available at time $t_n$ gives
\begin{equation}
\mathbf y_n^p=
\mathbf h_n^p(\mathbf X_{t_n}^p)
+\boldsymbol\varepsilon_n^p,
\end{equation}
with observation covariance
$\mathbf R_n^p=
\sigma_{\varepsilon,p}^2
\operatorname{diag}
\left(\tau_{k,n}\right)_{k\in\mathcal K_n}$,
where $\mathcal K_n$ is the set of available contracts and $\mathbf h_n^p(\mathbf x)=[F_k(t_n;\mathbf x)]_{k\in\mathcal K_n}$ maps the latent state to the vector of model-implied forward prices for these contracts. 

For the load model, observed DC power measurements are similarly related to the latent state through the seasonal component and polynomial observation mapping.

For given model parameters  $\boldsymbol\Theta_a=\{\boldsymbol\theta_a,\boldsymbol\rho_a,\boldsymbol\alpha_a,\beta_a,\boldsymbol\omega_a,\boldsymbol\chi_a,\sigma_{\varepsilon,a}\}$ 
the UKF uses the exact Gaussian state transition \eqref{eq:exact-step} and the nonlinear observation model \eqref{eq:observation} to estimate the latent states. It also provides the one-step-ahead innovation $\mathbf e_{n,a}=\mathbf y_{n,a}-\hat{\mathbf y}_{n|n-1,a}$ 
and its covariance $\boldsymbol\Sigma_{n,a}$, which are used to construct the approximate Gaussian log-likelihood:
\begin{equation}
\footnotesize
\log\mathcal L(\boldsymbol\Theta_a)=
-\frac12\sum_{n=1}^{N}
\left[
d_{n,a}\log(2\pi)+\log|\boldsymbol\Sigma_{n,a}|
+\mathbf e_{n,a}^\top\boldsymbol\Sigma_{n,a}^{-1}\mathbf e_{n,a}
\right]\label{eq:MLE}
\end{equation}

The model parameters $\boldsymbol\Theta_a$ 
are estimated offline by maximizing \eqref{eq:MLE} using a covariance matrix adaptation evolution strategy 
(CMA-ES) \cite{hansen2001cmaes}. \color{black} 

Offline calibration uses rolling historical data 
of two weeks 
to estimate the seasonal coefficients by least squares and the PDM parameters by approximate maximum likelihood. The generator matrix $\mathbf G_a$ and its decomposition are then precomputed. Online, the UKF updates the latent-state estimates every 15 minutes using new observations, and \eqref{eq:conditionalex} is evaluated at the filtered state estimate to update the price and load forecasts. Model parameters  $\boldsymbol\Theta$ remain fixed between recalibrations. The updated forecasts are used for BESS dispatch in Section III.

\section{Battery Dispatch Approach: Dynamic Programming}
\label{sec:battery}
The proposed DP framework optimizes BESS charging and discharging using the conditional price and load forecasts developed in Section II. Every 15 minutes, the forecasts and dispatch schedule are updated using new observations, with only the current decision implemented.

\subsection{Battery Model and Operating Constraints}

Let $\hat P_k=F_k(t_n)$ and $\hat L_k=L_k(t_n)$ denote the conditional price and load forecasts for delivery interval $k$, obtained at 
decision time $t_n$. The BESS state of charge is $b\in[b^{\min},b^{\max}]$, and  $\mathbf u_k=(u_{c,k},u_{d,k})$ 
denotes charging and discharging power, respectively. The battery state transition is
\begin{equation}
b_{k+1}=B(b_k,\mathbf u_k)
=b_k+\left(\eta_cu_{c,k}-\frac{u_{d,k}}{\eta_d}\right)\Delta t,
\end{equation}
where $\eta_c$ and $\eta_d$ are the charging and discharging efficiencies, and $\Delta t$ is the dispatch interval.

The feasible control set is
\begin{equation}
\footnotesize
\begin{split}
\mathcal U_k(b) = \Big\{ u :\;
0 \le u_c \le u_c^{\max},\;
0 \le u_d \le \min\big(u_d^{\max}, \hat L_k\big),\\\;
u_c u_d = 0,\;
b^{\min} \le B(b,\mathbf u) \le b^{\max} \Big\}.
\end{split}
\label{eq:feasible}
\end{equation}
These constraints enforce power and energy limits, 
prohibit simultaneous charging and discharging, and maintain the allowable battery energy range. Discharging is further limited by the forecast load to avoid scheduled power export.

Battery degradation is approximated by a linear throughput cost \cite{zhao2022quality}: 
$c_{\mathrm{wear}}(\mathbf u_k)
=\lambda(u_{c,k}+u_{d,k})\Delta t$, 
where $\lambda$ is the degradation cost per MWh of battery throughput. 

\subsection{Dynamic Programming Formulation}
Let $V_k(b)$ denote the minimum forecast-based cumulative cost from interval $k$ onward, given battery energy $b$. The Bellman recursion is
\begin{equation}
\small
\begin{split}
V_k(b) = \min_{\mathbf u_k\in \mathcal U_k(b)} \Big\{\,
  & \hat P_k\big(\hat L_k + u_{c,k} - u_{d,k}\big)\Delta t \\
  & +c_{\mathrm{wear}}(\mathbf u_k)
    + V_{k+1}\big( B(b,\mathbf u_k)\big) \Big\},
\end{split}
\label{eq:bellman}
\end{equation}
\normalsize
The terminal condition \(V_{K+1}(b)=W(b)\) approximates the cost-to-go beyond the optimization horizon. The terminal cost \(W(b)\) is estimated iteratively by averaging scenario-specific DP value functions until convergence, with \(W(b_{\min})=0\). 

At each 15-minute update, the DP is solved using the latest forecasts, and only the first optimal control action is implemented.

\subsection{Realized Cost and Performance Evaluation}

The realized electricity cost accounts for scheduled energy purchases, forecast errors, and battery degradation:   
\begin{equation}
\small
\begin{split}
C^{\mathrm{real}}=\sum_{k=1}^K\Big[&\pi^{\mathrm{tr}}_k\big(\hat L_k+u_{c,k}-u_{d,k}\big)\Delta t
+\pi^{\mathrm{dev}}_k\big(L_k-\hat L_k\big)\Delta t\\
&+c_{\mathrm{wear}}(\mathbf u_k)\Big].
\end{split}
\label{eq:realized-cost}
\end{equation}
\normalsize
where $\pi_k^{\mathrm{tr}}$ is the trading price, $\pi_k^{\mathrm{dev}}$ is the deviation settlement price, and $L_k$ is the realized load.

\begin{figure*}[t]
    \centering
    \includegraphics[width=0.8\textwidth]{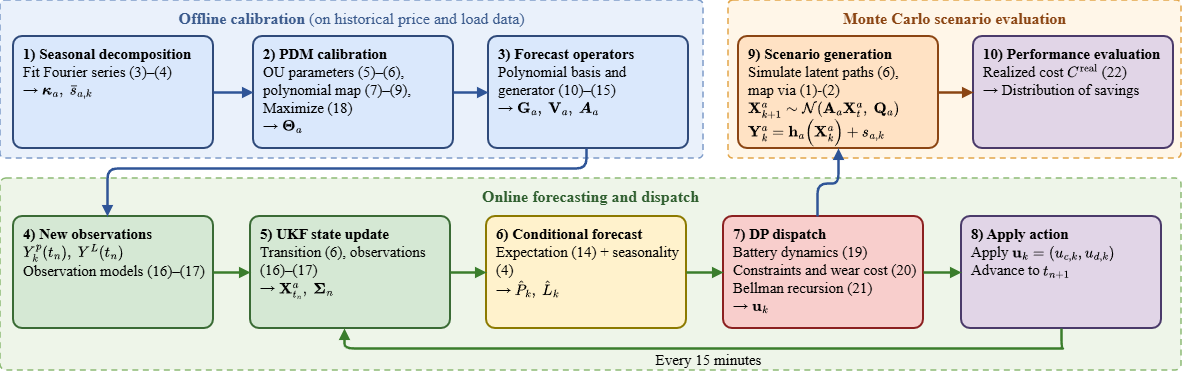}
    \caption{Workflow of the proposed PDM-based BESS dispatch framework, including offline model calibration, online UKF-based forecasting and DP dispatch, and Monte Carlo evaluation of cost savings.} 
    \label{fig:dispatch-overview}
\end{figure*}

As illustrated in Fig.~\ref{fig:dispatch-overview}, the calibrated PDMs generate stochastic price and load trajectories using the Gaussian transition \eqref{eq:exact-step}, polynomial observation mapping \eqref{eq:polynomial mapping}, and seasonal components \eqref{eq:window-ave}. These trajectories are used to estimate the distributions of realized cost savings based on \eqref{eq:realized-cost}.

\section{Validation and Results}
This section evaluates the PDM forecasting performance and the cost savings achieved by the proposed DP-based BESS dispatch strategy. 

\subsection{Polynomial Diffusion Model Validation}
\begin{figure}[t]
    \centering
    \includegraphics[width=\linewidth, height=2.6cm]{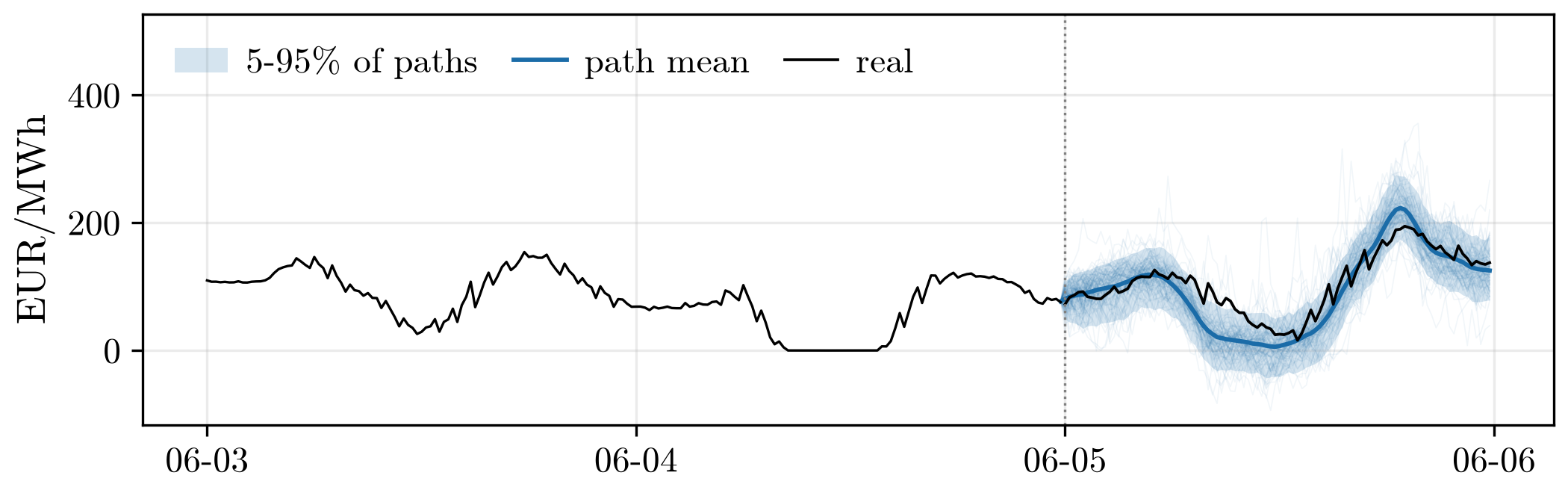}\\[-0.5ex]
    \includegraphics[width=\linewidth, height=2.6cm]{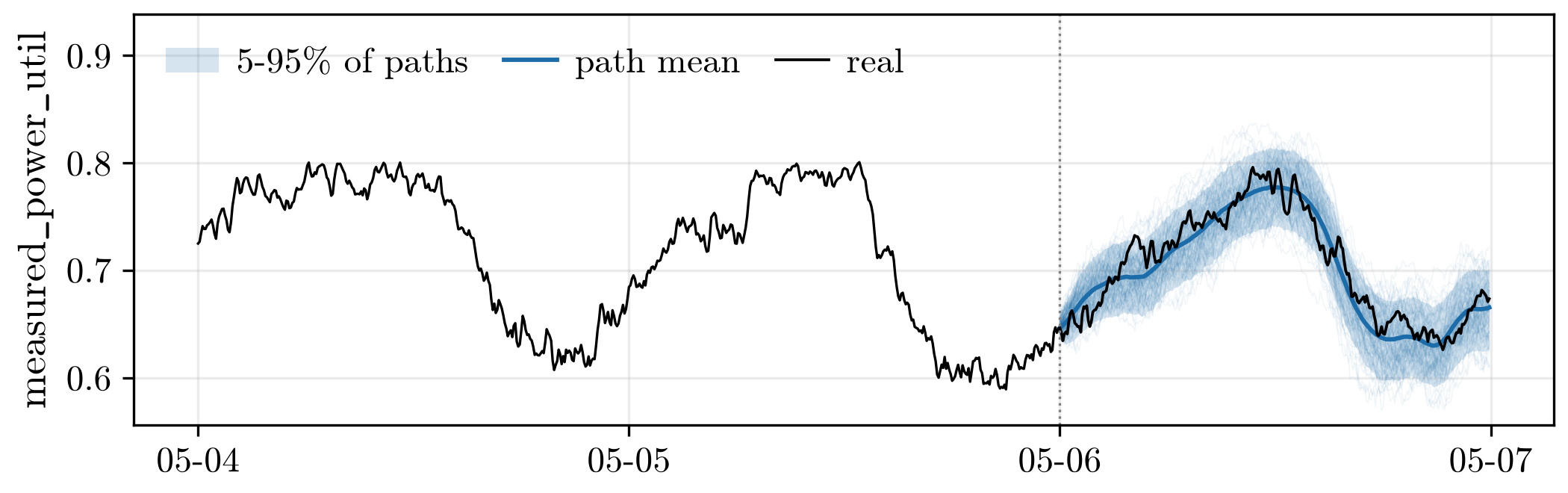}
    \vspace{-1ex}
    \caption{PDM-simulated and realized (a) electricity prices (top) and (b) data center loads (bottom) over a one-day out-of-sample period (1,000 simulations).} 
    \label{fig:validate}
    \vspace{-2ex}
\end{figure}

The price PDM is calibrated using historical German intraday electricity prices, while the load PDM uses publicly available Google DC power traces \cite{wilkes2020clusterdata}. Fig.~\ref{fig:validate} compares PDM-simulated trajectories with out-of-sample observations,  where both models capture the main temporal variations. 

Over 1,000 simulated trajectories during a five-day out-of-sample period, the PDMs yield mean electricity prices of €116.2/MWh (realized: €117.6/MWh) and normalized loads of 0.7171 (realized: 0.7143). The corresponding standard deviations are 67.9 versus 53.5 €/MWh for prices and 0.0527 versus 0.0489 for loads, indicating that the PDMs overestimate variability, particularly for electricity prices. 




\subsection{Results}
The proposed DP strategy is compared with a rule-based (RB) threshold strategy \cite{dabbagh2019shaving} through two studies: a benchmark with a 100-MW DC and a 12-MW/36-MWh BESS, similar to \cite{liu2026}, and a sensitivity analysis across BESS sizes. The system in \cite{liu2026} uses a degradation cost of approximately €40/MWh, with an assumed round-trip efficiency of about 90\%. The RB strategy we implement charges below €80/MWh, discharges above €150/MWh, and otherwise remains idle, subject to battery operating constraints. 

\begin{figure}
    \centering
    \includegraphics[width=0.8\linewidth]{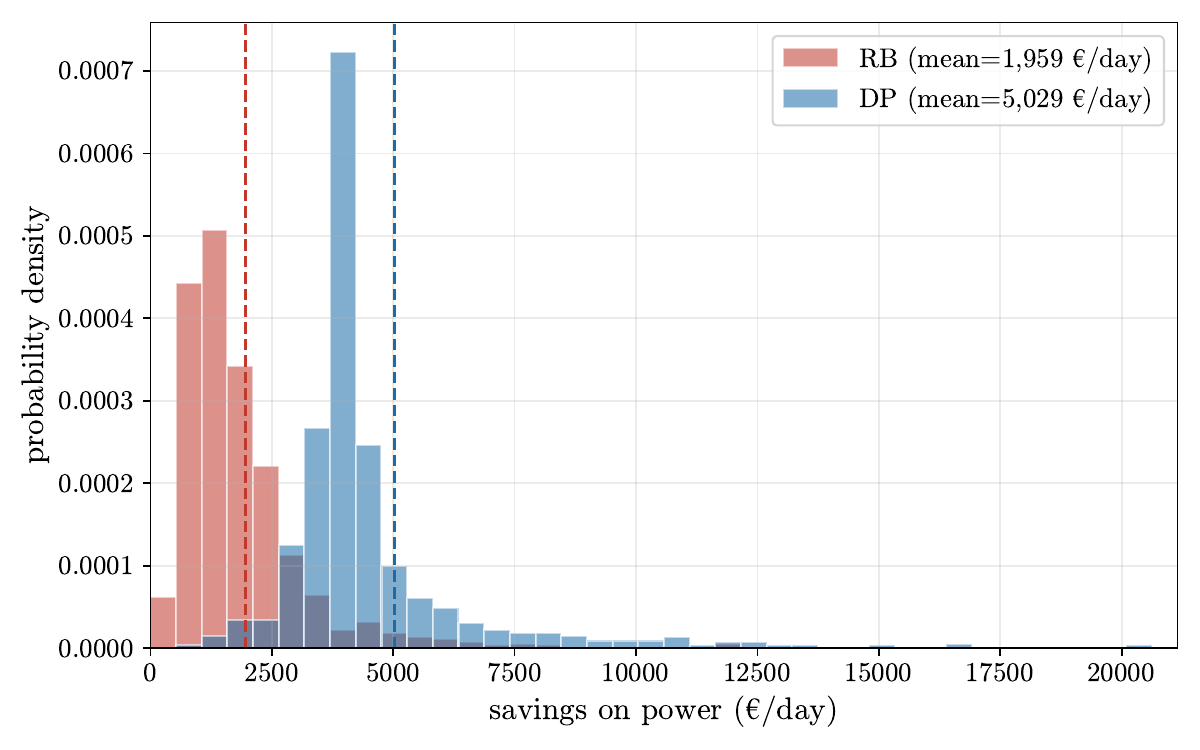}
    \caption{Distributions of BESS daily cost savings under DP and RB strategies 
    over 1,000 simulations.} 
    \label{fig:sdvsrb}
\end{figure}

Fig.~\ref{fig:sdvsrb} compares the distributions of daily cost savings over 1,000 simulations. DP achieves mean savings of €5,029/day, compared with €1,959/day for RB, representing a 2.57-fold improvement. These savings are comparable in magnitude to those reported in \cite{liu2026}, although differences in electricity markets and revenue streams prevent direct comparison. 

Table \ref{tab:sim-params} compares the two strategies across four BESS configurations for a 20-MW DC. DP consistently outperforms RB, achieving 2.53–2.74 times greater cost savings, with an aggregate savings ratio of 2.63.

\begin{figure}
    \centering
    \subfloat[Cost savings\label{fig:returns-vs-bat}]{%
        \includegraphics[width=0.48\linewidth]{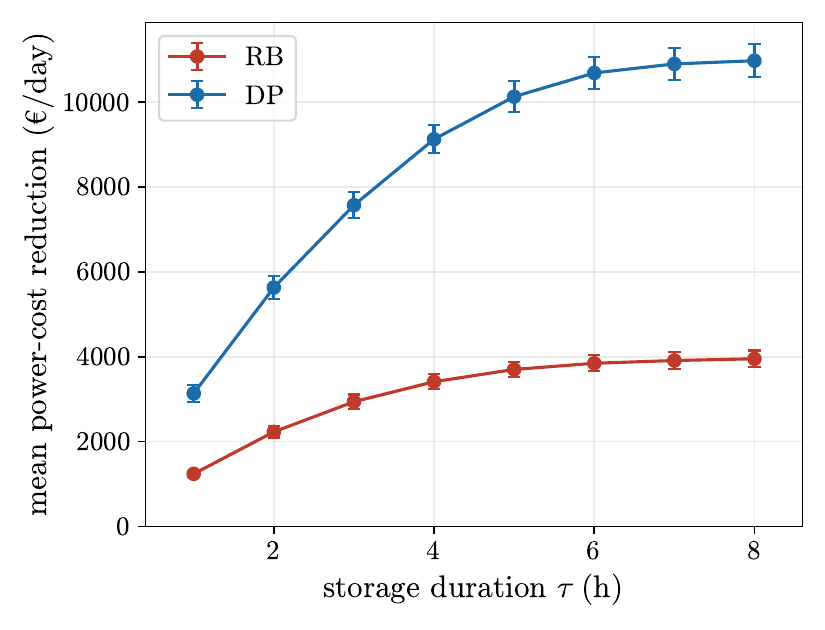}}%
    \hfill
    \subfloat[DP-to-RB ratios\label{fig:ratios-vs-bat}]{%
        \includegraphics[width=0.48\linewidth]{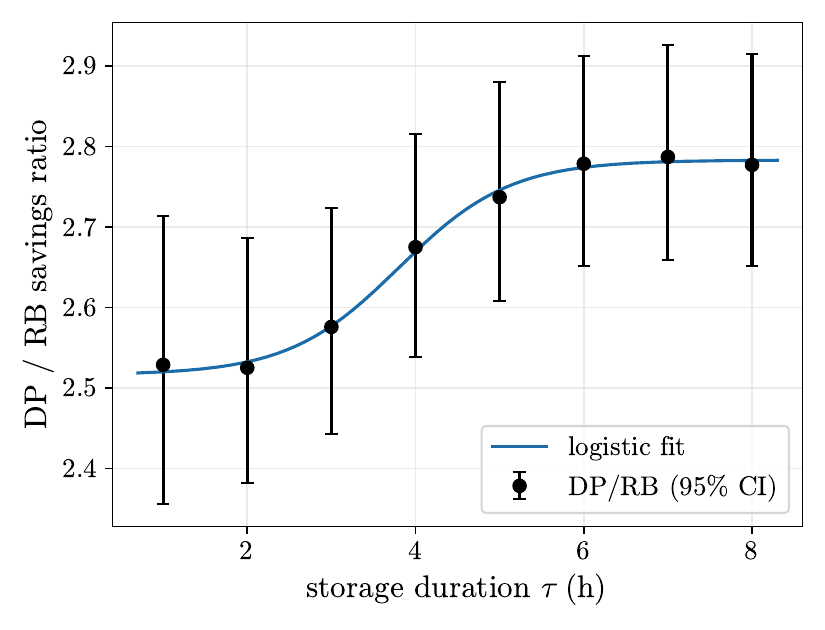}}
    \caption{Effects of BESS storage duration ($\tau$) on (a) mean daily cost savings and (b) DP-to-RB savings ratio.} 
    \label{fig:ratios}
\end{figure}

Fig.~\ref{fig:ratios}  further examines the impact of battery storage duration, defined as \(\tau=E_{\max}/P_{\max}\). As shown in Fig.~\ref{fig:ratios}(a), daily cost savings increase with storage duration for both strategies, with diminishing marginal gains at larger capacities. Fig.~\ref{fig:ratios}(b) shows that the DP-to-RB savings ratio generally increases before leveling off. This indicates that forecast-based dispatch can better exploit the flexibility of longer-duration storage, whereas RB relies on fixed price thresholds.

\begin{table}[t]
\centering
\caption{Mean daily cost savings (€/day) for DP vs.\ RB across battery configurations (20-MW DC, 1{,}000 simulations).}
\label{tab:sim-params}
\begin{tabular}{cccc}
\toprule
\multicolumn{2}{c}{Battery} & \multicolumn{2}{c}{Cost savings (€/day)} \\
\cmidrule(lr){1-2} \cmidrule(lr){3-4}
$P_{\max}$ (MW) & $E_{\max}$ (MWh) & RB & DP \\
\midrule
20 & 20  & 1{,}241 & 3{,}137 \\
10 & 30  & 1{,}632 & 4{,}191 \\
20 & 60  & 2{,}940 & 7{,}573 \\
20 & 100 & 3{,}701 & 10{,}130 \\
\bottomrule
\end{tabular}
\end{table}


\section{Conclusion and Future Work}
This paper developed a PDM-based forecasting and DP framework for cost-effective BESS dispatch in DCs. By extending PDMs to DC loads, the framework enables efficient conditional forecasting of both electricity prices and power demand. Simulation results demonstrate that the proposed DP strategy achieves an aggregate cost savings ratio of 2.63 relative to RB dispatch across different BESS configurations. Future work will incorporate higher-order predictive moments to enable risk-aware BESS operation.
\bibliographystyle{IEEEtran}
\bibliography{ref} 

\end{document}